\documentstyle[12pt]{article}

\ifx\undefined\psfig\else \fi

\edef\psfigRestoreAt{\catcode`@=\number\catcode`@\relax}
\catcode`\@=11\relax
\newwrite\@unused
\def\typeout#1{{\let\protect\string\immediate\write\@unused{#1}}}
\def\figurepath{./}

\def\@nnil{\@nil}
\def\@empty{}
\def\@psdonoop#1\@@#2#3{}
\def\@psdo#1:=#2\do#3{\edef\@psdotmp{#2}\ifx\@psdotmp\@empty \else
    \expandafter\@psdoloop#2,\@nil,\@nil\@@#1{#3}\fi}
\def\@psdoloop#1,#2,#3\@@#4#5{\def#4{#1}\ifx #4\@nnil \else
       #5\def#4{#2}\ifx #4\@nnil \else#5\@ipsdoloop #3\@@#4{#5}\fi\fi}
\def\@ipsdoloop#1,#2\@@#3#4{\def#3{#1}\ifx #3\@nnil 
       \let\@nextwhile=\@psdonoop \else
      #4\relax\let\@nextwhile=\@ipsdoloop\fi\@nextwhile#2\@@#3{#4}}
\def\@tpsdo#1:=#2\do#3{\xdef\@psdotmp{#2}\ifx\@psdotmp\@empty \else
    \@tpsdoloop#2\@nil\@nil\@@#1{#3}\fi}
\def\@tpsdoloop#1#2\@@#3#4{\def#3{#1}\ifx #3\@nnil 
       \let\@nextwhile=\@psdonoop \else
      #4\relax\let\@nextwhile=\@tpsdoloop\fi\@nextwhile#2\@@#3{#4}}
\newread\ps@stream
\newif\ifnot@eof       
\newif\if@noisy        
\newif\if@atend        
\newif\if@psfile       
{\catcode`\%=12\global\gdef\epsf@start{
\def\epsf@PS{PS}
\def\epsf@getbb#1{%
\openin\ps@stream=#1
\ifeof\ps@stream\typeout{Error, File #1 not found}\else
   {\not@eoftrue \chardef\other=12
    \def\do##1{\catcode`##1=\other}\dospecials \catcode`\ =10
    \loop
       \if@psfile
	  \read\ps@stream to \epsf@fileline
       \else{
	  \obeyspaces
          \read\ps@stream to \epsf@tmp\global\let\epsf@fileline\epsf@tmp}
       \fi
       \ifeof\ps@stream\not@eoffalse\else
       \if@psfile\else
       \expandafter\epsf@test\epsf@fileline:. \\%
       \fi
          \expandafter\epsf@aux\epsf@fileline:. \\%
       \fi
   \ifnot@eof\repeat
   }\closein\ps@stream\fi}%
\long\def\epsf@test#1#2#3:#4\\{\def\epsf@testit{#1#2}
			\ifx\epsf@testit\epsf@start\else
\typeout{Warning! File does not start with `\epsf@start'.  It may not be a PostScript file.}
			\fi
			\@psfiletrue} 
{\catcode`\%=12\global\let\epsf@percent=
\long\def\epsf@aux#1#2:#3\\{\ifx#1\epsf@percent
   \def\epsf@testit{#2}\ifx\epsf@testit\epsf@bblit
	\@atendfalse
        \epsf@atend #3 . \\%
	\if@atend	
	   \if@verbose{
		\typeout{psfig: found `(atend)'; continuing search}
	   }\fi
        \else
        \epsf@grab #3 . . . \\%
        \not@eoffalse
        \global\no@bbfalse
        \fi
   \fi\fi}%
\def\epsf@grab #1 #2 #3 #4 #5\\{%
   \global\def\epsf@llx{#1}\ifx\epsf@llx\empty
      \epsf@grab #2 #3 #4 #5 .\\\else
   \global\def\epsf@lly{#2}%
   \global\def\epsf@urx{#3}\global\def\epsf@ury{#4}\fi}%
\def\epsf@atendlit{(atend)} 
\def\epsf@atend #1 #2 #3\\{%
   \def\epsf@tmp{#1}\ifx\epsf@tmp\empty
      \epsf@atend #2 #3 .\\\else
   \ifx\epsf@tmp\epsf@atendlit\@atendtrue\fi\fi}

\chardef\letter = 11
\chardef\other = 12

\newif \ifdebug 
\newif\ifc@mpute 
\c@mputetrue 

\let\then = \relax
\def\r@dian{pt }
\let\r@dians = \r@dian
\let\dimensionless@nit = \r@dian
\let\dimensionless@nits = \dimensionless@nit
\def\internal@nit{sp }
\let\internal@nits = \internal@nit
\newif\ifstillc@nverging
\def \Mess@ge #1{\ifdebug \then \message {#1} \fi}

{ 
	\catcode `\@ = \letter
	\gdef \nodimen {\expandafter \n@dimen \the \dimen}
	\gdef \term #1 #2 #3%
	       {\edef \t@ {\the #1}
		\edef \t@@ {\expandafter \n@dimen \the #2\r@dian}%
		\t@rm {\t@} {\t@@} {#3}%
	       }
	\gdef \t@rm #1 #2 #3%
	       {{%
		\count 0 = 0
		\dimen 0 = 1 \dimensionless@nit
		\dimen 2 = #2\relax
		\Mess@ge {Calculating term #1 of \nodimen 2}%
		\loop
		\ifnum	\count 0 < #1
		\then	\advance \count 0 by 1
			\Mess@ge {Iteration \the \count 0 \space}%
			\Multiply \dimen 0 by {\dimen 2}%
			\Mess@ge {After multiplication, term = \nodimen 0}%
			\Divide \dimen 0 by {\count 0}%
			\Mess@ge {After division, term = \nodimen 0}%
		\repeat
		\Mess@ge {Final value for term #1 of 
				\nodimen 2 \space is \nodimen 0}%
		\xdef \Term {#3 = \nodimen 0 \r@dians}%
		\aftergroup \Term
	       }}
	\catcode `\p = \other
	\catcode `\t = \other
	\gdef \n@dimen #1pt{#1} 
}

\def \Divide #1by #2{\divide #1 by #2} 

\def \Multiply #1by #2
       {{
	\count 0 = #1\relax
	\count 2 = #2\relax
	\count 4 = 65536
	\Mess@ge {Before scaling, count 0 = \the \count 0 \space and
			count 2 = \the \count 2}%
	\ifnum	\count 0 > 32767 
	\then	\divide \count 0 by 4
		\divide \count 4 by 4
	\else	\ifnum	\count 0 < -32767
		\then	\divide \count 0 by 4
			\divide \count 4 by 4
		\else
		\fi
	\fi
	\ifnum	\count 2 > 32767 
	\then	\divide \count 2 by 4
		\divide \count 4 by 4
	\else	\ifnum	\count 2 < -32767
		\then	\divide \count 2 by 4
			\divide \count 4 by 4
		\else
		\fi
	\fi
	\multiply \count 0 by \count 2
	\divide \count 0 by \count 4
	\xdef \product {#1 = \the \count 0 \internal@nits}%
	\aftergroup \product
       }}

\def\r@duce{\ifdim\dimen0 > 90\r@dian \then   
		\multiply\dimen0 by -1
		\advance\dimen0 by 180\r@dian
		\r@duce
	    \else \ifdim\dimen0 < -90\r@dian \then  
		\advance\dimen0 by 360\r@dian
		\r@duce
		\fi
	    \fi}

\def\Sine#1%
       {{%
	\dimen 0 = #1 \r@dian
	\r@duce
	\ifdim\dimen0 = -90\r@dian \then
	   \dimen4 = -1\r@dian
	   \c@mputefalse
	\fi
	\ifdim\dimen0 = 90\r@dian \then
	   \dimen4 = 1\r@dian
	   \c@mputefalse
	\fi
	\ifdim\dimen0 = 0\r@dian \then
	   \dimen4 = 0\r@dian
	   \c@mputefalse
	\fi
	\ifc@mpute \then
		\divide\dimen0 by 180
		\dimen0=3.141592654\dimen0
		\dimen 2 = 3.1415926535897963\r@dian 
		\divide\dimen 2 by 2 
		\Mess@ge {Sin: calculating Sin of \nodimen 0}%
		\count 0 = 1 
		\dimen 2 = 1 \r@dian 
		\dimen 4 = 0 \r@dian 
		\loop
			\ifnum	\dimen 2 = 0 
			\then	\stillc@nvergingfalse 
			\else	\stillc@nvergingtrue
			\fi
			\ifstillc@nverging 
			\then	\term {\count 0} {\dimen 0} {\dimen 2}%
				\advance \count 0 by 2
				\count 2 = \count 0
				\divide \count 2 by 2
				\ifodd	\count 2 
				\then	\advance \dimen 4 by \dimen 2
				\else	\advance \dimen 4 by -\dimen 2
				\fi
		\repeat
	\fi		
			\xdef \sine {\nodimen 4}%
       }}

\def\Cosine#1{\ifx\sine\UnDefined\edef\Savesine{\relax}\else
		             \edef\Savesine{\sine}\fi
	{\dimen0=#1\r@dian\advance\dimen0 by 90\r@dian
	 \Sine{\nodimen 0}
	 \xdef\cosine{\sine}
	 \xdef\sine{\Savesine}}}	      

\def\psdraft{
	\def\@psdraft{0}
}
\def\psfull{
	\def\@psdraft{100}
}

\psfull

\newif\if@draftbox
\def\psnodraftbox{
	\@draftboxfalse
}
\def\psdraftbox{
	\@draftboxtrue
}
\@draftboxtrue

\newif\if@prologfile
\newif\if@postlogfile
\def\pssilent{
	\@noisyfalse
}
\def\psnoisy{
	\@noisytrue
}
\psnoisy
\newif\if@bbllx
\newif\if@bblly
\newif\if@bburx
\newif\if@bbury
\newif\if@height
\newif\if@width
\newif\if@rheight
\newif\if@rwidth
\newif\if@angle
\newif\if@clip
\newif\if@verbose
\def\@p@@sclip#1{\@cliptrue}

\def\@p@@sfile#1{\def\@p@sfile{null}%
	        \openin1=#1
		\ifeof1\closein1%
		       \openin1=\figurepath#1
			\ifeof1\typeout{Error, File #1 not found}
			   \if@bbllx\if@bblly\if@bburx\if@bbury
			      \def\@p@sfile{#1}%
			   \fi\fi\fi\fi
			\else\closein1
			    \edef\@p@sfile{\figurepath#1}%
                        \fi%
		 \else\closein1%
		       \def\@p@sfile{#1}%
		 \fi}
\def\@p@@sfigure#1{\def\@p@sfile{null}%
	        \openin1=#1
		\ifeof1\closein1%
		       \openin1=\figurepath#1
			\ifeof1\typeout{Error, File #1 not found}
			   \if@bbllx\if@bblly\if@bburx\if@bbury
			      \def\@p@sfile{#1}%
			   \fi\fi\fi\fi
			\else\closein1
			    \def\@p@sfile{\figurepath#1}%
                        \fi%
		 \else\closein1%
		       \def\@p@sfile{#1}%
		 \fi}

\def\@p@@sbbllx#1{
		\@bbllxtrue
		\dimen100=#1
		\edef\@p@sbbllx{\number\dimen100}
}
\def\@p@@sbblly#1{
		\@bbllytrue
		\dimen100=#1
		\edef\@p@sbblly{\number\dimen100}
}
\def\@p@@sbburx#1{
		\@bburxtrue
		\dimen100=#1
		\edef\@p@sbburx{\number\dimen100}
}
\def\@p@@sbbury#1{
		\@bburytrue
		\dimen100=#1
		\edef\@p@sbbury{\number\dimen100}
}
\def\@p@@sheight#1{
		\@heighttrue
		\dimen100=#1
   		\edef\@p@sheight{\number\dimen100}
}
\def\@p@@swidth#1{
		\@widthtrue
		\dimen100=#1
		\edef\@p@swidth{\number\dimen100}
}
\def\@p@@srheight#1{
		\@rheighttrue
		\dimen100=#1
		\edef\@p@srheight{\number\dimen100}
}
\def\@p@@srwidth#1{
		\@rwidthtrue
		\dimen100=#1
		\edef\@p@srwidth{\number\dimen100}
}
\def\@p@@sangle#1{
		\@angletrue
		\edef\@p@sangle{#1} 
}
\def\@p@@ssilent#1{ 
		\@verbosefalse
}
\def\@p@@sprolog#1{\@prologfiletrue\def\@prologfileval{#1}}
\def\@p@@spostlog#1{\@postlogfiletrue\def\@postlogfileval{#1}}
\def\@cs@name#1{\csname #1\endcsname}
\def\@setparms#1=#2,{\@cs@name{@p@@s#1}{#2}}
\def\ps@init@parms{
		\@bbllxfalse \@bbllyfalse
		\@bburxfalse \@bburyfalse
		\@heightfalse \@widthfalse
		\@rheightfalse \@rwidthfalse
		\def\@p@sbbllx{}\def\@p@sbblly{}
		\def\@p@sbburx{}\def\@p@sbbury{}
		\def\@p@sheight{}\def\@p@swidth{}
		\def\@p@srheight{}\def\@p@srwidth{}
		\def\@p@sangle{0}
		\def\@p@sfile{}
		\def\@p@scost{10}
		\def\@sc{}
		\@prologfilefalse
		\@postlogfilefalse
		\@clipfalse
		\if@noisy
			\@verbosetrue
		\else
			\@verbosefalse
		\fi
}
\def\parse@ps@parms#1{
	 	\@psdo\@psfiga:=#1\do
		   {\expandafter\@setparms\@psfiga,}}
\newif\ifno@bb
\def\bb@missing{
	\if@verbose{
		\typeout{psfig: searching \@p@sfile \space  for bounding box}
	}\fi
	\no@bbtrue
	\epsf@getbb{\@p@sfile}
        \ifno@bb \else \bb@cull\epsf@llx\epsf@lly\epsf@urx\epsf@ury\fi
}	
\def\bb@cull#1#2#3#4{
	\dimen100=#1 bp\edef\@p@sbbllx{\number\dimen100}
	\dimen100=#2 bp\edef\@p@sbblly{\number\dimen100}
	\dimen100=#3 bp\edef\@p@sbburx{\number\dimen100}
	\dimen100=#4 bp\edef\@p@sbbury{\number\dimen100}
	\no@bbfalse
}
\newdimen\p@intvaluex
\newdimen\p@intvaluey
\def\rotate@#1#2{{\dimen0=#1 sp\dimen1=#2 sp
		  \global\p@intvaluex=\cosine\dimen0
		  \dimen3=\sine\dimen1
		  \global\advance\p@intvaluex by -\dimen3
		  \global\p@intvaluey=\sine\dimen0
		  \dimen3=\cosine\dimen1
		  \global\advance\p@intvaluey by \dimen3
		  }}
\def\compute@bb{
		\no@bbfalse
		\if@bbllx \else \no@bbtrue \fi
		\if@bblly \else \no@bbtrue \fi
		\if@bburx \else \no@bbtrue \fi
		\if@bbury \else \no@bbtrue \fi
		\ifno@bb \bb@missing \fi
		\ifno@bb \typeout{FATAL ERROR: no bb supplied or found}
			\no-bb-error
		\fi
		\if@angle 
			\Sine{\@p@sangle}\Cosine{\@p@sangle}
	        	{\dimen100=\maxdimen\xdef\r@p@sbbllx{\number\dimen100}
					    \xdef\r@p@sbblly{\number\dimen100}
			                    \xdef\r@p@sbburx{-\number\dimen100}
					    \xdef\r@p@sbbury{-\number\dimen100}}
                        \def\minmaxtest{
			   \ifnum\number\p@intvaluex<\r@p@sbbllx
			      \xdef\r@p@sbbllx{\number\p@intvaluex}\fi
			   \ifnum\number\p@intvaluex>\r@p@sbburx
			      \xdef\r@p@sbburx{\number\p@intvaluex}\fi
			   \ifnum\number\p@intvaluey<\r@p@sbblly
			      \xdef\r@p@sbblly{\number\p@intvaluey}\fi
			   \ifnum\number\p@intvaluey>\r@p@sbbury
			      \xdef\r@p@sbbury{\number\p@intvaluey}\fi
			   }
			\rotate@{\@p@sbbllx}{\@p@sbblly}
			\minmaxtest
			\rotate@{\@p@sbbllx}{\@p@sbbury}
			\minmaxtest
			\rotate@{\@p@sbburx}{\@p@sbblly}
			\minmaxtest
			\rotate@{\@p@sbburx}{\@p@sbbury}
			\minmaxtest
			\edef\@p@sbbllx{\r@p@sbbllx}\edef\@p@sbblly{\r@p@sbblly}
			\edef\@p@sbburx{\r@p@sbburx}\edef\@p@sbbury{\r@p@sbbury}
		\fi
		\count203=\@p@sbburx
		\count204=\@p@sbbury
		\advance\count203 by -\@p@sbbllx
		\advance\count204 by -\@p@sbblly
		\edef\@bbw{\number\count203}
		\edef\@bbh{\number\count204}
}
\def\in@hundreds#1#2#3{\count240=#2 \count241=#3
		     \count100=\count240	
		     \divide\count100 by \count241
		     \count101=\count100
		     \multiply\count101 by \count241
		     \advance\count240 by -\count101
		     \multiply\count240 by 10
		     \count101=\count240	
		     \divide\count101 by \count241
		     \count102=\count101
		     \multiply\count102 by \count241
		     \advance\count240 by -\count102
		     \multiply\count240 by 10
		     \count102=\count240	
		     \divide\count102 by \count241
		     \count200=#1\count205=0
		     \count201=\count200
			\multiply\count201 by \count100
		 	\advance\count205 by \count201
		     \count201=\count200
			\divide\count201 by 10
			\multiply\count201 by \count101
			\advance\count205 by \count201
		     \count201=\count200
			\divide\count201 by 100
			\multiply\count201 by \count102
			\advance\count205 by \count201
		     \edef\@result{\number\count205}
}
\def\compute@wfromh{
		\in@hundreds{\@p@sheight}{\@bbw}{\@bbh}
		\edef\@p@swidth{\@result}
}
\def\compute@hfromw{
	        \in@hundreds{\@p@swidth}{\@bbh}{\@bbw}
		\edef\@p@sheight{\@result}
}
\def\compute@handw{
		\if@height 
			\if@width
			\else
				\compute@wfromh
			\fi
		\else 
			\if@width
				\compute@hfromw
			\else
				\edef\@p@sheight{\@bbh}
				\edef\@p@swidth{\@bbw}
			\fi
		\fi
}
\def\compute@resv{
		\if@rheight \else \edef\@p@srheight{\@p@sheight} \fi
		\if@rwidth \else \edef\@p@srwidth{\@p@swidth} \fi
}
\def\compute@sizes{
	\compute@bb
	\compute@handw
	\compute@resv
}
\def\psfig#1{\vbox {
	\ps@init@parms
	\parse@ps@parms{#1}
	\compute@sizes
	\ifnum\@p@scost<\@psdraft{
		\if@verbose{
			\typeout{psfig: including \@p@sfile \space }
		}\fi
		\special{ps::[begin] 	\@p@swidth \space \@p@sheight \space
				\@p@sbbllx \space \@p@sbblly \space
				\@p@sbburx \space \@p@sbbury \space
				startTexFig \space }
		\if@angle
			\special {ps:: \@p@sangle \space rotate \space} 
		\fi
		\if@clip{
			\if@verbose{
				\typeout{(clip)}
			}\fi
			\special{ps:: doclip \space }
		}\fi
		\if@prologfile
		    \special{ps: plotfile \@prologfileval \space } \fi
		\special{ps: plotfile \@p@sfile \space }
		\if@postlogfile
		    \special{ps: plotfile \@postlogfileval \space } \fi
		\special{ps::[end] endTexFig \space }
		\vbox to \@p@srheight true sp{
			\hbox to \@p@srwidth true sp{
				\hss
			}
		\vss
		}
	}\else{
		\if@draftbox{		
			\hbox{\fbox{\vbox to \@p@srheight true sp{
			\vss
			\hbox to \@p@srwidth true sp{ \hss \@p@sfile \hss }
			\vss
			}}}
		}\else{
			\vbox to \@p@srheight true sp{
			\vss
			\hbox to \@p@srwidth true sp{\hss}
			\vss
			}
		}\fi

	}\fi
}}
\def\psglobal{\typeout{psfig: PSGLOBAL is OBSOLETE; use psprint -m instead}}
\psfigRestoreAt

\def\ctr#1{{\it #1}\\\vspace{10pt}}
 
\begin{document}
\pagestyle{empty}

\vspace{2mm}

\hfill DESY-96-139\\[1cm]
\begin{center}

{\Large{\bf {The Presampler for the Forward and Rear Calorimeter in the ZEUS Detector}}}        

\vspace{1cm}

A.Bamberger$^4$, A.Bornheim$^1$, J.Crittenden$^1$, H.-J.Grabosch$^3$,
M.Grothe$^1$,
L.Hervas$^7$, E.Hilger$^1$, U.Holm$^5$, D.Horstmann$^5$,
V.Kaufmann$^4$, A.Kharchilava$^3$, U.K\"{o}tz$^2$,
D.Kummerow$^5$, U.Mallik$^6$, A.Meyer$^3$, M.Nowoczyn$^5$, R.Ossowski$^1$,
S.Schlenstedt$^3$, H.Tiecke$^8$,
W.Verkerke$^8$, J.Vossebeld$^8$, M.Vreeswijk$^8$, S.M.Wang$^6$, J.Wu$^6$

\vspace{1cm}

$^1$ Physikalisches Institut der Universit\"{a}t Bonn, Bonn, Germany \\
$^2$ Deutsches Elektronen-Synchrotron DESY, Hamburg, Germany \\
$^3$ DESY-IfH Zeuthen, Zeuthen, Germany \\
$^4$ Fakult\"{a}t f\"{u}r Physik der Universit\"{a}t Freiburg i.Br., Freiburg i.Br., Germany \\
$^5$ Hamburg University, II. Institute of Exp. Physics, Hamburg, Germany  \\
$^6$ University of Iowa Physics and Astronomy Dept, Iowa City, USA \\
$^7$ Univer. Autonoma Madrid, Depto de Fisica Teorica, Madrid, Spain \\
$^8$ NIKHEF and University of Amsterdam, Netherlands \\

\vspace{2cm}

\normalsize { \bf Abstract}

\end{center}

The ZEUS detector at HERA has been supplemented with a presampler detector
in front of the forward and rear calorimeters. It consists of a segmented
scintillator array read out with wavelength-shifting fibers. We discuss its design,
construction and performance. Test beam data obtained
with a prototype presampler and the ZEUS prototype
calorimeter demonstrate the main function
of this detector, i.e. the correction for the energy lost by
an electron interacting in inactive material in front of the calorimeter.

\vspace{2cm}

\newpage
\pagestyle{plain}

\section {Introduction}

The material situated between the electron-proton interaction point
and the front face of the uranium-scintillator calorimeter within the ZEUS
detector at HERA \cite{statusreport} leads to a degradation of the calorimetric energy 
measurement of the particles produced in the interaction.
We have constructed a presampler detector, now installed
directly in front of the forward and rear ZEUS calorimeter sections,
with the goal of measuring this energy degradation on an event--by--event basis.
The detector consists of a layer of scintillator tiles;
wavelength-shifting fibers, embedded in the scintillator, guide the
scintillation light to photomultipliers.
Particles which shower in the material in front of the presampler lead to
an increased particle multiplicity which is measured by the presampler. The
combined information from the presampler and the calorimeter allows 
an event-by-event measurement of the energy loss in front of the
calorimeter and thus allows to recover the energy scale and energy resolution of
the ZEUS calorimeter.  
We describe the production and performance of the optical
components, the assembly of the presampler detector, the readout system
and the calibration system.
We summarize results on the response of the individual tiles to
cosmic rays and results of a test of a prototype presampler
in combination with the ZEUS calorimeter prototype in a CERN test beam.

\section {Scintillator/fiber combination}

The segmentation of the presampler matches that of the
ZEUS calorimeter \cite{calpap} hadronic sections, 20~x~20 cm$^2$. 
The scintillation light is read out by wavelength-shifting fibers
embedded in the scintillator and transported by clear 
fibers to             
a photomultiplier. Since the calibration of the                      
presampler is performed with minimum--ionising particles (MIP) we
require a                  
photoelectron yield of at least 5 photoelectrons
per MIP at the photocathode             
of the photomultiplier~(PMT).

We have investigated several combinations of scintillator material,
fiber material and fiber layout to optimize light yield and response
uniformity, resulting in the following choice:

\begin{itemize} 
\item  scintillator material SCSN38 (Kuraray Co. Ltd), 5~mm thick     
 with dimension 203~x~198.5 mm$^2$;  tiles are cut and diamond-polished.
 Six grooves, parallel to the long edge, are machined in the tiles with an ordinary saw blade
\footnote {The blade has a diameter of 95~mm and is 1.2~mm thick; the revolution
speed was 2000-2500 turns/min and progressed with
about 10cm/min; as coolant we used a mixture
of water and detergent.}; the grooves are 1.2~mm wide, 1.5~mm deep and equally
spaced over the surface.

\item six fibers read out one tile; they are glued
\footnote {NE 581 Optical Cement with hardener} in the grooves in the
  scintillator.
\item  WLS fiber material Y11 (Kuraray Co. Ltd) double clad, 1~mm diameter, 23~cm long with
       a sputtered Al mirror at one end.
\item  a 1 mm diameter double clad transparent polystyrene fiber
       DCLG (Kuraray Co. Ltd) guides the light to the photomultiplier;
        the fiber is 3~m long and is glued to the WLS fiber.
\item  wrapping of the tile in Tyvek paper (quality Q173-D, DuPont) paper.                                                     
\end{itemize}                                                                   

\subsection {Fiber production.}

The faces of the fibers were smoothed in steps with sandpaper of
various granularities down
to a grain size of a few microns. For this
purpose a mounting support was designed which held 50 fibers
individually. In this manner the cladding was prevented from cracking and
the resulting face was perpendicular to the fibre-axis.
One face of each WLS-fiber was also polished and then aluminized
by vacuum evaporation
in order to prevent light loss through this end.\\
For an optimal joint between the WLS-fiber and the clear readout-fiber
a gluing procedure was chosen \footnote{BICRON BC600 Optical Cement}.
 Steel tubes, 
with an inner (outer) diameter of 1.1 (1.2)~mm, served for good matching of the
fibers and reinforcement of the joint. Special care
was taken to get a tight joint by applying gentle pressure during hardening.

All fibers were tested for light transmission using a monitored UV-lamp.
The fibers with lowest transmission were discarded (10\%) resulting in a spread
of  12\% for the remaining fibers.

\subsection{Light yield measurement}

\begin{figure}[tbp]
\centerline{\psfig{figure=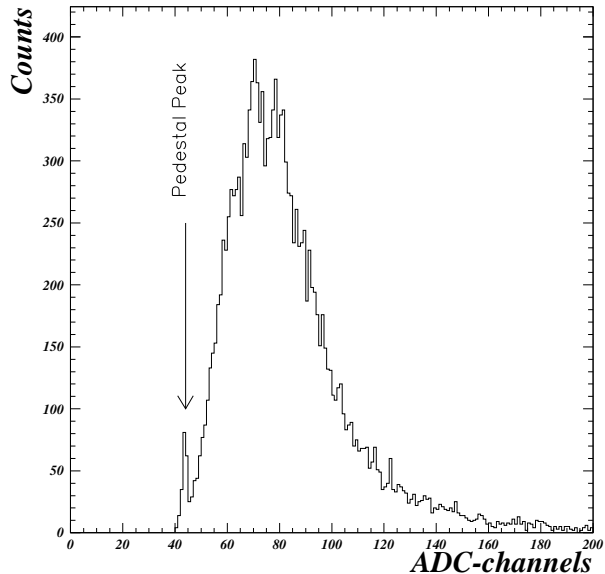,bbllx=0pt,bblly=140pt,bburx=550pt,bbury=650pt,height=8cm}}
   \caption{\it Response of a scintillator tile to cosmic muons. The peak 
 around the pedestal value of 45 ADC counts is due to false triggers 
and zero response due to photostatistics.}            
   \label{cos.ps}
\end{figure}

The absolute light yield for minimum--ionising particles was measured       
in a cosmic ray telescope with an effective area of 12~x~12 cm$^{2}$.
The PMT pulse was integrated by a LeCroy 2249A ADC,            
triggered by a threefold coincidence of the signals from the cosmic           
trigger\footnote{The absolute trigger efficiency was determined to
be 98.9\%}. The ADC gate length was 100 ns.
Figure<~\ref{cos.ps} shows the response to cosmic ray particles
for the final tile/fiber/PMT layout. 

The ratio of triggers with no signal above pedestal              
recorded in the ADC to the total number of triggers is equal to
(0.37$\pm$0.03\%), taking into account the trigger efficiency.
From this we obtain
an absolute light yield of 5.6$\pm$.1 photoelectrons.

\subsection{Uniformity measurement}

\begin{figure}[tbp]
\centerline{\psfig{figure=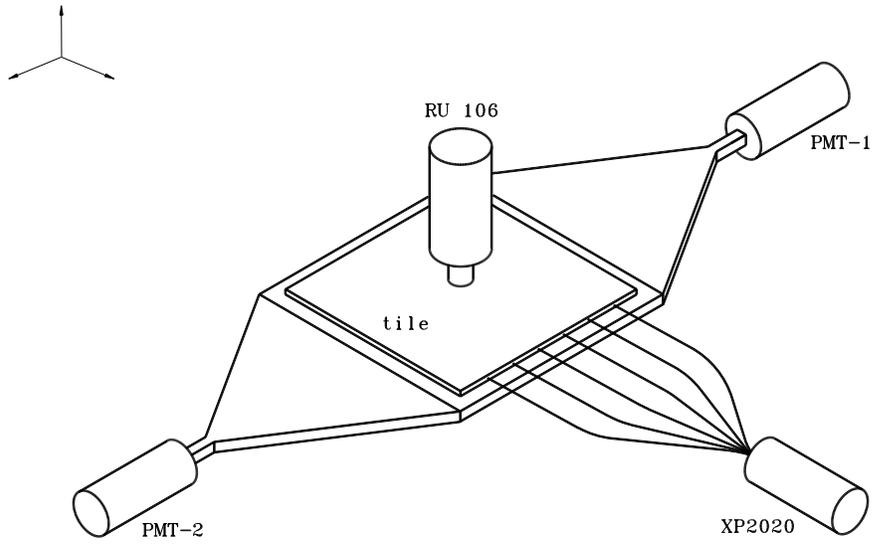,bbllx=130pt,bblly=55pt,bburx=450pt,bbury=600pt,height=8cm,angle=270}}
   \caption{\it Scanning table uniformity measurement}
   \label{tab.ps}
\end{figure}

The tile uniformity was measured with  a                   
collimated $^{106}$Ru source, a computer-controlled x-y scanning table
as scintillator support and an XP2020 photomultiplier.
The ADC was triggered by a coincidence            
of two photomultiplier signals reading out a large trigger counter              
located underneath the tile to be tested (figure~\ref{tab.ps}). To ensure that the             
$^{106}$Ru source simulates minimum--ionising particles,
we measured the light yield of the tile also    
with the cosmic telescope. The pulse height spectra of                 
cosmic muons and electrons from the $^{106}$Ru source are very similar  
in shape, with peak values equal within 5\%.
A 3--mm--diameter collimator was used to measure the response
over a 20~x~20 cm$^{2}$ tile in steps of 2~mm in x and y. 
The response is uniform within 5\%.

\begin{figure}[tbp]
\centerline{\psfig{figure=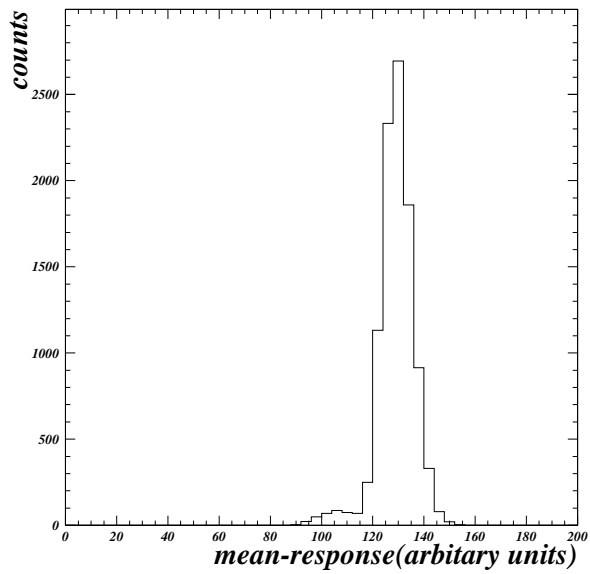,bbllx=15pt,bblly=150pt,bburx=550pt,bbury=670pt,height=8cm}}
   \caption{\it Spread of response over a tile. The response is uniform within 5\%.}
   \label{mean.ps}
\end{figure}

The distribution of the mean values            
of the measurements for all positions is shown in figure~\ref{mean.ps}.
The slight bump below 120 channels is an effect occurring near the edge of the tile
where some particles leave the tile before crossing its full thickness.

\section {Mechanical layout}

\subsection {Tile assembly}

The scintillator tiles are assembled in cassettes, made of 0.4~mm
thick stainless steel; they are 20~cm wide and vary in length,                
containing between 1 and 10 tiles. In order for
the readout fibers to clear the neighboring scintillator tile
the 5~mm thick tiles are raised
at the readout end, supported by a 2.5~mm thick rohacell strip. 
The total thickness of the cassette is
$\simeq$ 11~mm and represents about 5~\% of a radiation length (1.2~\% for the
scintillator and 4~\% for the stainless steel).
The length of readout fibers is $\simeq$ 300~cm.
The six readout fibers of one tile are glued together in a connector which
is fixed to the PMT housing. In addition to the six fibers
a seventh clear fiber is added to guide the light from a laser/LED system
to the PMT.

\subsection {Detector assembly}

The forward and rear
calorimeters are split in two halves, such that they can be withdrawn
from the beam pipe region during injection of the electrons and protons in HERA.
A group of 19~cassettes, which covers one half of each calorimeter face,
is glued on a 2~mm thick aluminium plate (0.02 radiation lengths) of 2~x~4~m$^2$.

\begin{figure}[tbp]
\centerline{
\psfig{figure=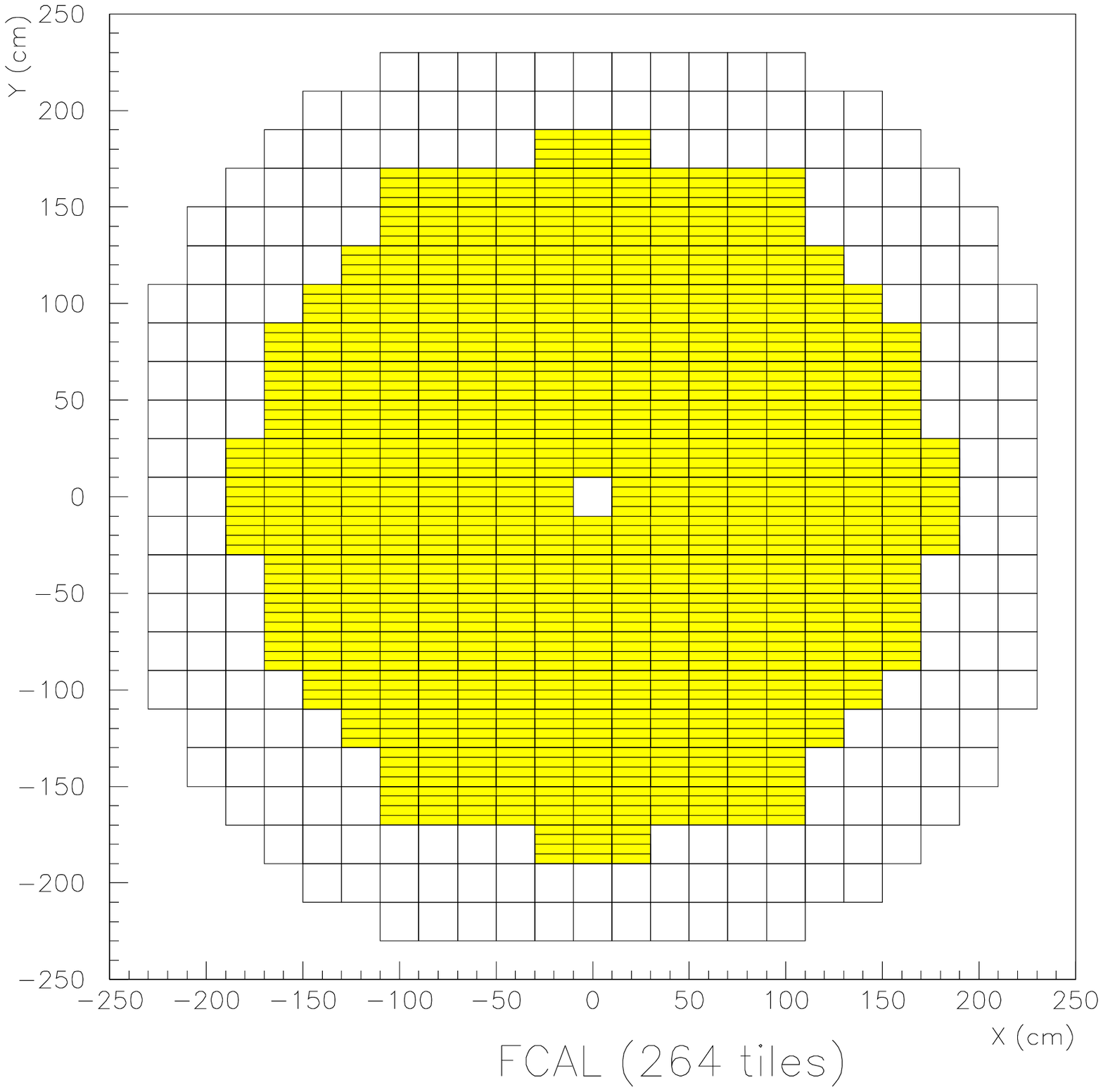,bbllx=0pt,bblly=100pt,bburx=580pt,bbury=680pt,height=8cm}
\psfig{figure=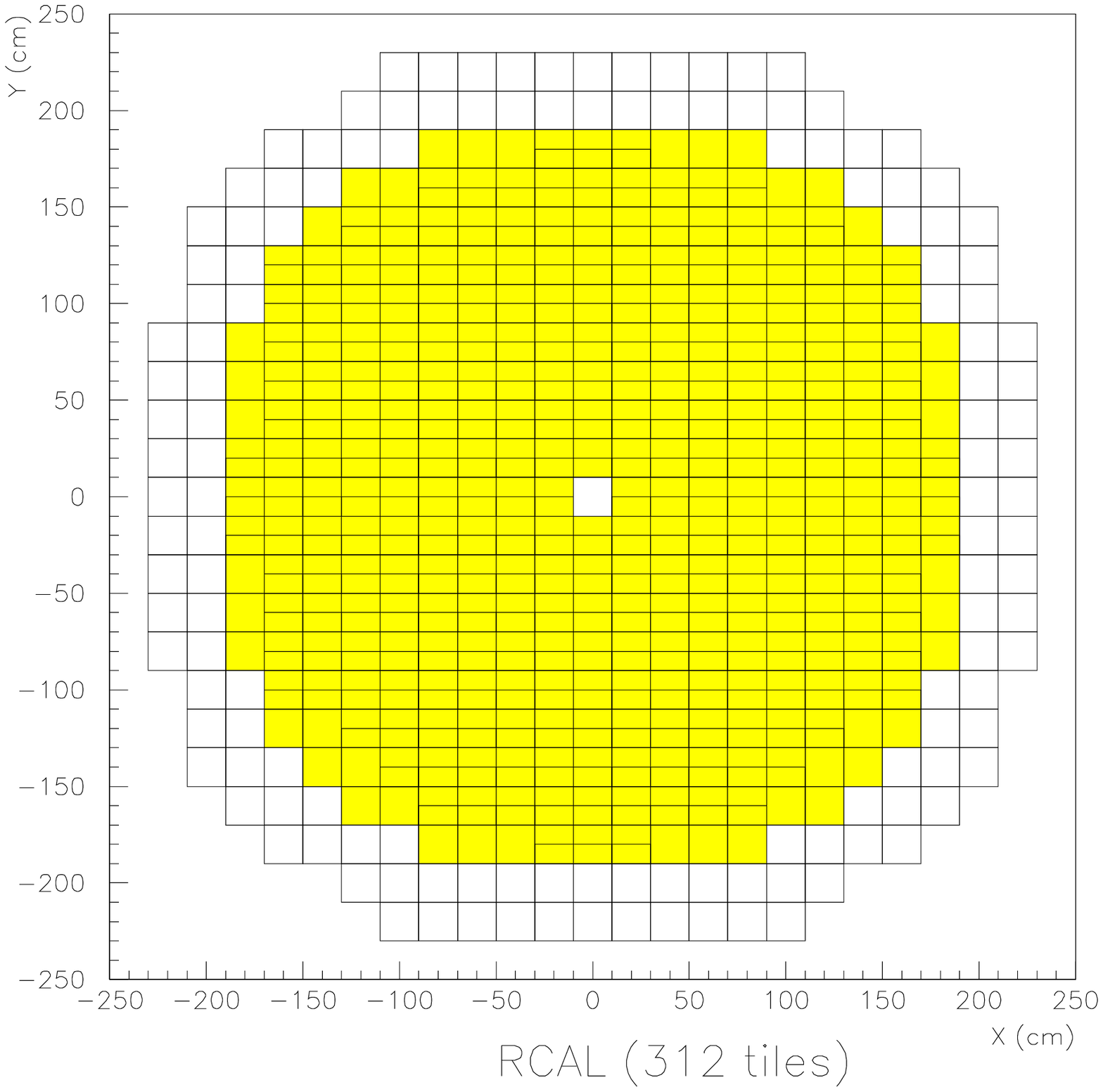,bbllx=0pt,bblly=100pt,bburx=580pt,bbury=680pt,height=8cm}}
   \caption{\it 
Front view of the forward (FCAL) and rear calorimeter (RCAL). The 20~x~20~cm$^{2}$
white square in the center corresponds to the hole for the beam pipe. The coverage
of the presampler is indicated by the shaded region.}
   \label{layout_presam}
\end{figure}

Figure~\ref{layout_presam} shows the coverage of the calorimeter by the presampler.
Shown is the segmentation of the electromagnetic sections, which is
finer in the region not shadowed from the nominal interaction point by the
barrel calorimeter. The 20~x~20~cm$^{2}$ towers covered by the presampler tiles are shaded.

A 2.5~mm diameter tube is glued over the full length on the outside of each
cassette, positioned at the center of the tiles.
The tube guides a radioactive source for calibrating the light output
of the individual tiles and the gain of the PMT channels (see section 7.2).

\section {Photomultiplier tests}
\subsection{Performance specifications}

Due to the limited space available in the ZEUS detector it was decided 
to use multi\-chan\-nel PMT's. 
The magnetic field amounts to a few hundred Gauss in the area where
the PMT's are located and therefore adequate shielding is needed.
Since we measure pulse heights, the crosstalk between adjacent
channels in the tube is required to be less than 5\%. 
Another requirement is
the size of the photocathode for a single channel, which must match the readout fibers
of one scintillator tile.

The Hamamatsu R4760 16-channel photomultiplier has been extensively tested
for our application (see also \cite{R476ref}).    This is a 4~x~4 multichannel
PMT with a front face of 70~mm  diameter. Each of the 16 channels
has a 10 stage dynode chain, but they all share the same
voltage divider. The diameter of the photocathode for each channel 
is 8~mm. Our PMT's fulfill the following requirements:

\begin{itemize}
\item cathode sensitivity $>$ 45 $\mu$A/lm
\item minimum gain at 1000 V: 1~x~10$^6$
\item gain spread between channels, within one PMT assembly, less than a factor of 3
\end{itemize}

The crosstalk has been measured to be less than 3\%. 

\subsection{HV supply and linearity measurement}
A high voltage system based on the use of a
Cockcroft--Walton generator has been developed \cite{henk_hv}.
 Power dissipation is negligible compared to that
of a resistive voltage divider. The system can
be safely operated because of the low voltage input and provides
a protection against high currents (light leaks);
the maximum anode current is 100$\mu$A. 
The HV units consist of a microprocessor board and the voltage
multiplier boards. The microprocessor performs the HV setting and monitoring.

\begin{figure}[tbp]
\centerline{\psfig{figure=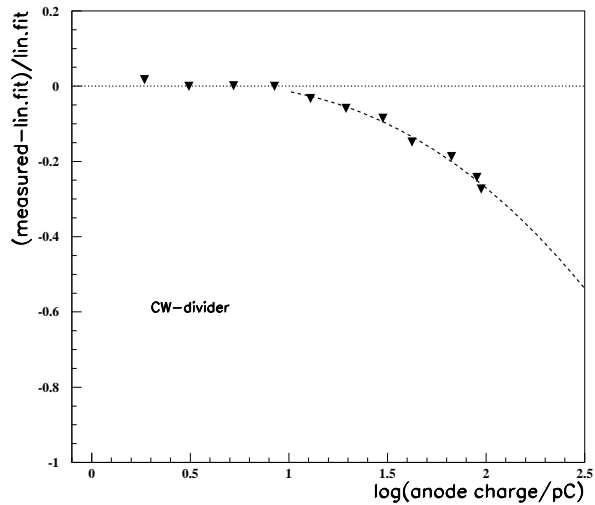,bbllx=0pt,bblly=150pt,bburx=550pt,bbury=670pt,height=8cm}}
   \caption{\it Results on the PMT nonlinearity.
     The line shows the linear fit
     through the first four points. The dashed line represents a
     polynomial fit through the last four points.}
   \label{pmlinea}
\end{figure}

The R4760 PMT operates in the range 800-1200V. 
Tests showed that the optimum linearity is obtained (at the
cost of a slightly lower gain) if the dynode voltage differences
are distributed in the proportions 2:1:1:1:1:1:2:2:4:3, starting at the cathode.
Since a pulse height measurement is required,
good knowledge of the relation between input charge and output signal
is necessary. For the linearity measurement we used
an LED and a linear neutral density filter. As an example we show in
figure~\ref{pmlinea} the deviation from linear behaviour versus
anode charge for a HV setting of 1100V. The dotted line shows 
the linear fit through the first four points, the dashed line represents 
a polynomial fit through the last four points.
 At an anode charge of
25 pC the measured values are about five percent lower 
than expected for linear behaviour. These 
values vary strongly from channel to channel and from PMT to PMT.

\section {Readout system}

The readout system is a copy  of the existing ZEUS calorimeter
readout system with some minor modifications \cite{caldwell}.
The PMT pulses are amplified and shaped by a pulse shaper circuit
mounted at the detector. The shaped pulse is sampled
every 96 ns (the bunch crossing 
rate of the HERA storage ring) and stored in a switched capacitor analog pipeline.
After receipt of a trigger from the ZEUS detector, eight samples are
transferred from the pipeline to an analog buffer and multiplexed to ADCs.
The data are sent to a location outside the detector where the
digitisation and signal processing takes place.
The modifications consist of upgraded versions of the
shaping/amplifier~\cite{iris} and the digital signal processor and a different
mechanical layout of the analog front end cards.

\section {Results from cosmic ray measurements}

\begin{figure}[htbp]
\centerline{\psfig{figure=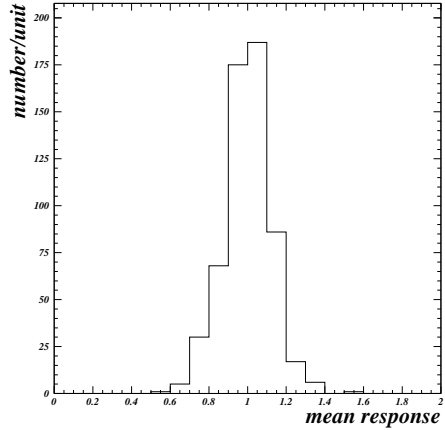,bbllx=45pt,bblly=160pt,bburx=530pt,bbury=650pt,height=5.7cm}}
   \caption{\it Average pulse height for cosmic muons normalized to one after having
 corrected for the gain differences between individual PMT channels.}
   \label{fig_tst}
\end{figure}

  A cosmic ray test was performed to measure the light yield of
the 576 tiles ( 264 FCAL and 312 RCAL tiles) assembled in 76 cassettes. 
The trigger system consists of eight cosmic ray telescopes. 
Each of these consists of two scintillator pads, 20~cm apart,
with an area of 12~x~12~cm$^{2}$ which give, together
with a third scintillator counter of 240~x~12~cm$^{2}$, a trigger system with a
three-fold  coincidence. The efficiency of each single telescope was
better than 98\%. The readout PMT was a 16-channel R4760 as used in
the final presampler design. The setup allows the measurement of 16 channels
simultaneously ( e.g. two cassettes with eight tiles each). The
trigger rate of the complete setup was about 50 counts/min
with about 6 counts/min for a single telescope. An example of the
cosmic-test measurement is given in figure~\ref{cos.ps}.
To compare the light yield of different tiles, all 16 channels of the
PMT were calibrated with a reference tile to correct for differences in
quantum efficiency (QE) and gain between the 16 PMT channels.
Figure~\ref{fig_tst} shows
the mean value for all 576 tiles normalized to one of them.
From the RMS value of the distribution we 
conclude that the responses of all tiles are equal to within 12\%.

\begin{figure}[htbp]
\centerline{\psfig{figure=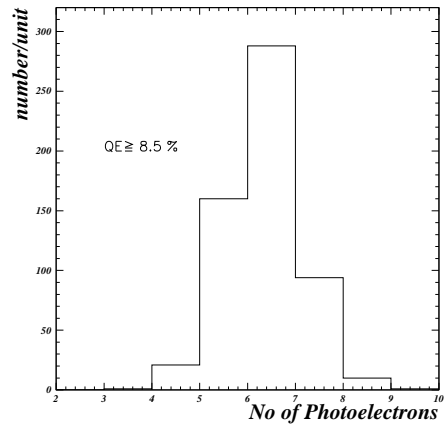,bbllx=45pt,bblly=150pt,bburx=530pt,bbury=650pt,height=5.7cm}}
   \caption{\it Average number of photoelectrons per tile per MIP}
   \label{fig_npe}
\end{figure}

Figure~\ref{fig_npe} shows the mean number of photoelectrons for each tile
assuming a QE of 8.5\% which is the minimum value accepted for the
presampler PMT's (the mean QE for all channels is 11.6\%).

\vskip 3cm

\section {Calibration tools}
\subsection{Minimum--ionising particles in situ}

During the operation of ZEUS, halo muons 
and charged hadrons are used to determine the response to single particles for each
individual channel. The high voltage setting common to the sixteen pixels of
one R4760 PMT is chosen such that the pixel with the least gain has
an average response to minimum--ionising particles which is a factor of ten
greater than the RMS noise level of the analog signal--processing front--end
electronics (0.05 pC). The pixel--to--pixel gain variation of about a factor of three
within one PMT results in a similar variation in the saturation levels.
The in situ calibration for the 1995 running period achieved a precision
of better than 5\% per tile.

\subsection{The radioactive source system}

We use an LED/laser system to monitor the gains of the PMT's and
a source system to monitor the combined response of tile, fiber and PMT.
The response to a $^{60}$Co source provides a relative calibration and quality
control of the individual channels of the presampler.
The source scans take place during shutdowns of HERA and provide
information on the long term behaviour of the light output of the
combination of scintillator and wavelength-shifting fiber.

\begin{figure}[htbp]
\centerline{\psfig{figure=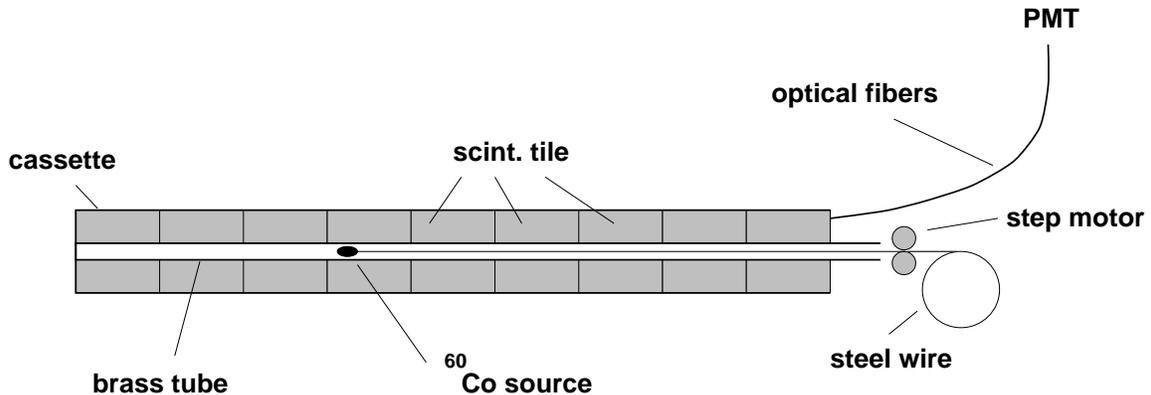,bbllx=203pt,bblly=0pt,bburx=0pt,bbury=608pt,angle=270,width=15cm}}
   \caption{\it Example of a cassette with 9 scintillating tiles and source tube
 glued on top, connected  to source scanning system.}
   \label{source_layout}
\end{figure}

Brass tubes with 2.5 mm outer diameter and 0.2 mm wall thickness
run over the full length of the cassette, positioned
in the middle (figure~\ref{source_layout}).
They guide the pointlike (0.8~mm diameter, 1~mm length, 74 MBq) source.
The source is driven in 2~mm steps via a 1.2~mm diameter
steel wire by a stepper motor~\cite{source_ref} 
controlled by a PC. The PMT currents are integrated
with a time constant of 24 ms and read into a 16-channel 12-bit ADC card.

\begin{figure}[htbp]
\centerline{\psfig{figure=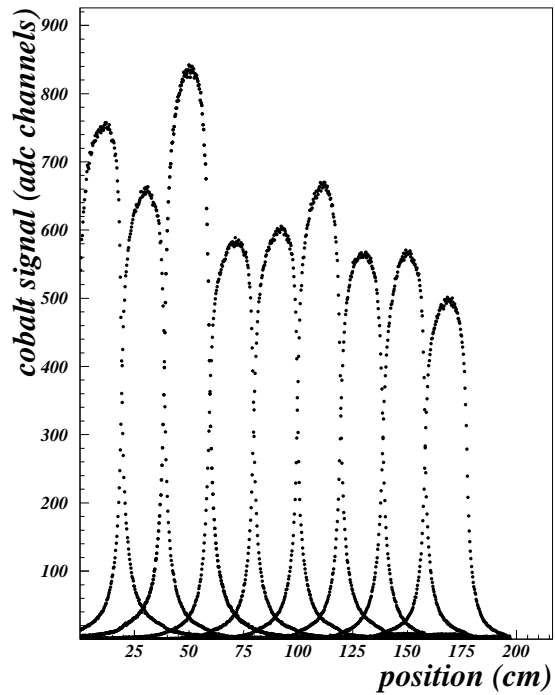,width=8cm,height=10cm}}
   \caption{\it  Responses of the scintillating tiles within one cassette
to a $^{60}$Co source. The step width is approximately 2 mm.}
   \label{x3}
\end{figure}

Figure~\ref{x3} shows as an example the superposition of the 
responses of the tiles within one cassette as a function of
the location of the source. The different heights of the 
maxima are mainly due to the different gains of
the 16 channels of the R4760 PMT, all supplied with the same
high voltage. The $^{60}$Co source is not collimated, as can be seen 
in the shape of the individual peaks.

\clearpage

\section {Beam test results}

The influence of material in front of the ZEUS
calorimeter on its energy measurement
has been studied previously in several test beam runs with
the ZEUS forward calorimeter (FCAL) prototype \cite{prot}.
The corrections to the calorimetric measurements that can be derived from
presampling measurements have been studied in subsequent test periods
for both hadrons and electrons~\cite{marcel}. In the following we summarize
the most recent results obtained for electrons with the final presampler
design~\cite{adi}.

\begin{figure}[tbp]
\centerline{\psfig{figure=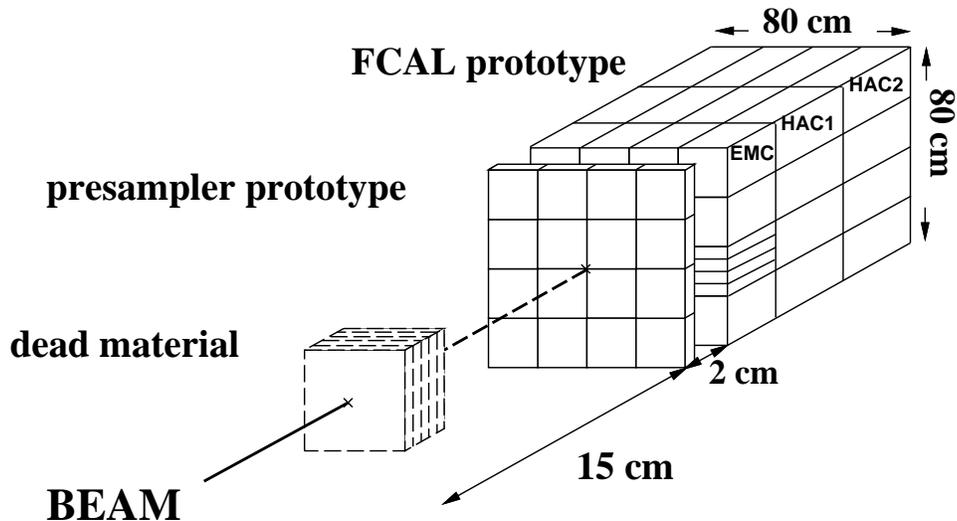,bbllx=30pt,bblly=440pt,bburx=565pt,bbury=700pt,height=6cm}}
  \caption{\it Experimental setup of the FCAL prototype and presampler in CERN
test beam. The presampler is mounted directly on the FCAL frontplate.}
   \label{preprot}
\end{figure}

\subsection{Overview}
\normalsize

The presampler prototype
consists of an array of 4 x 4 scintillator tiles covering
an area of 80 x 80 cm$^2$.  As shown in figure~\ref{preprot}
it is positioned directly in front of the ZEUS FCAL
prototype which has the same lateral size.
The depth of the calorimeter is 7 interaction lengths \cite{prot}.
Beam tests were  performed in the X5 test beam of the CERN SPS West Area.
The prototype presampler detector is read out via an R4760 multichannel
photomultiplier using the Cockcroft--Walton HV system. Furthermore, the
final readout electronics was used for both the presampler and the FCAL
prototype modules. 

The uranium radioactivity was used to
set the relative gains of the calorimeter 
phototubes and 15 GeV electrons served to set
the energy scale.
Muons were used
to calibrate the presampler.  The combined response of the
presampler and calorimeter was determined for electrons 
in the energy range from 3-50 GeV. The amount of
material installed in front of the presampler varied between
0 and 4 radiation lengths (X$_0$) of aluminium. 
During these studies, the position of both
calorimeter and presampler relative to the beam was fixed.
A delay wire chamber allowed the determination
of the impact point of the beam particles with an accuracy
of 0.5~mm.

Most of the data were recorded with a 
defocussed beam about 10 cm in diameter, facilitating studies of uniformity and
position dependence of the energy correction algorithms. 

\subsection{ The uniformity of the presampler response to muons }

\begin{figure}[hb]
\begin{picture}(400,170)(0,0)
%
\put(0,18) {\psfig{figure=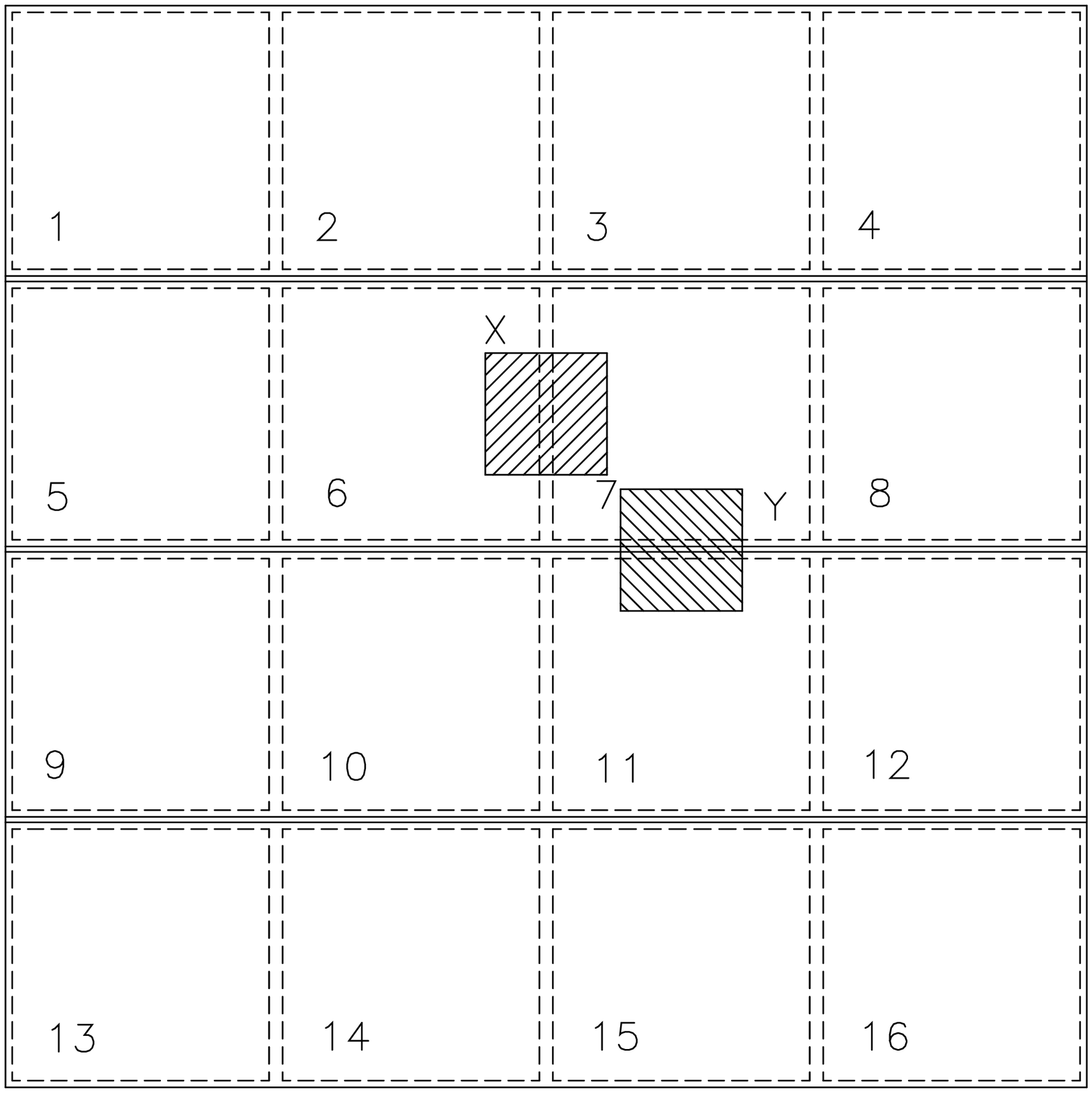,bbllx=6pt,bblly=40pt,bburx=580pt,bbury=630pt,height=4.7cm}}
\put(130,0) {\psfig{figure=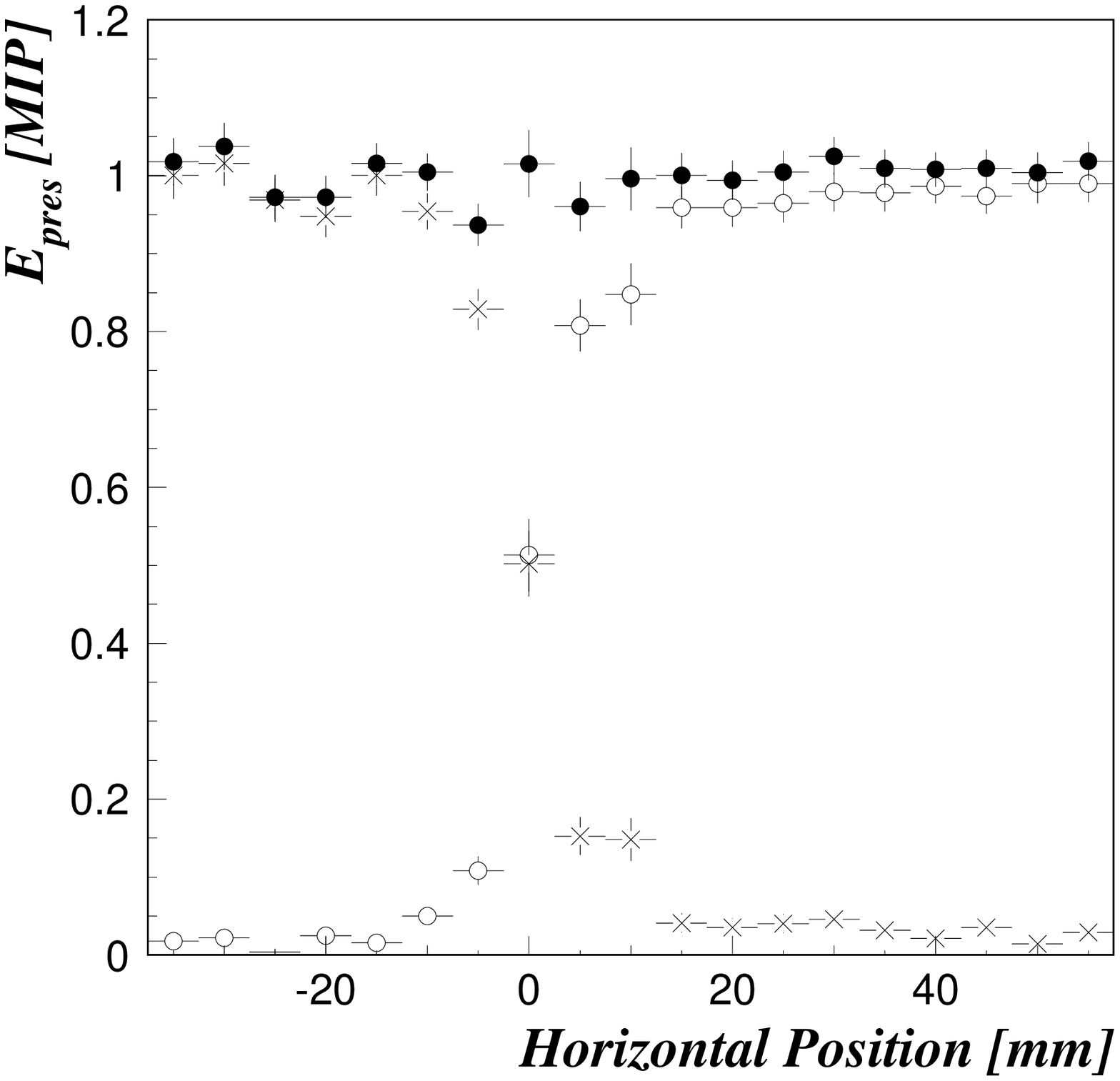,height=6cm}}
\put(285,0){\psfig{figure=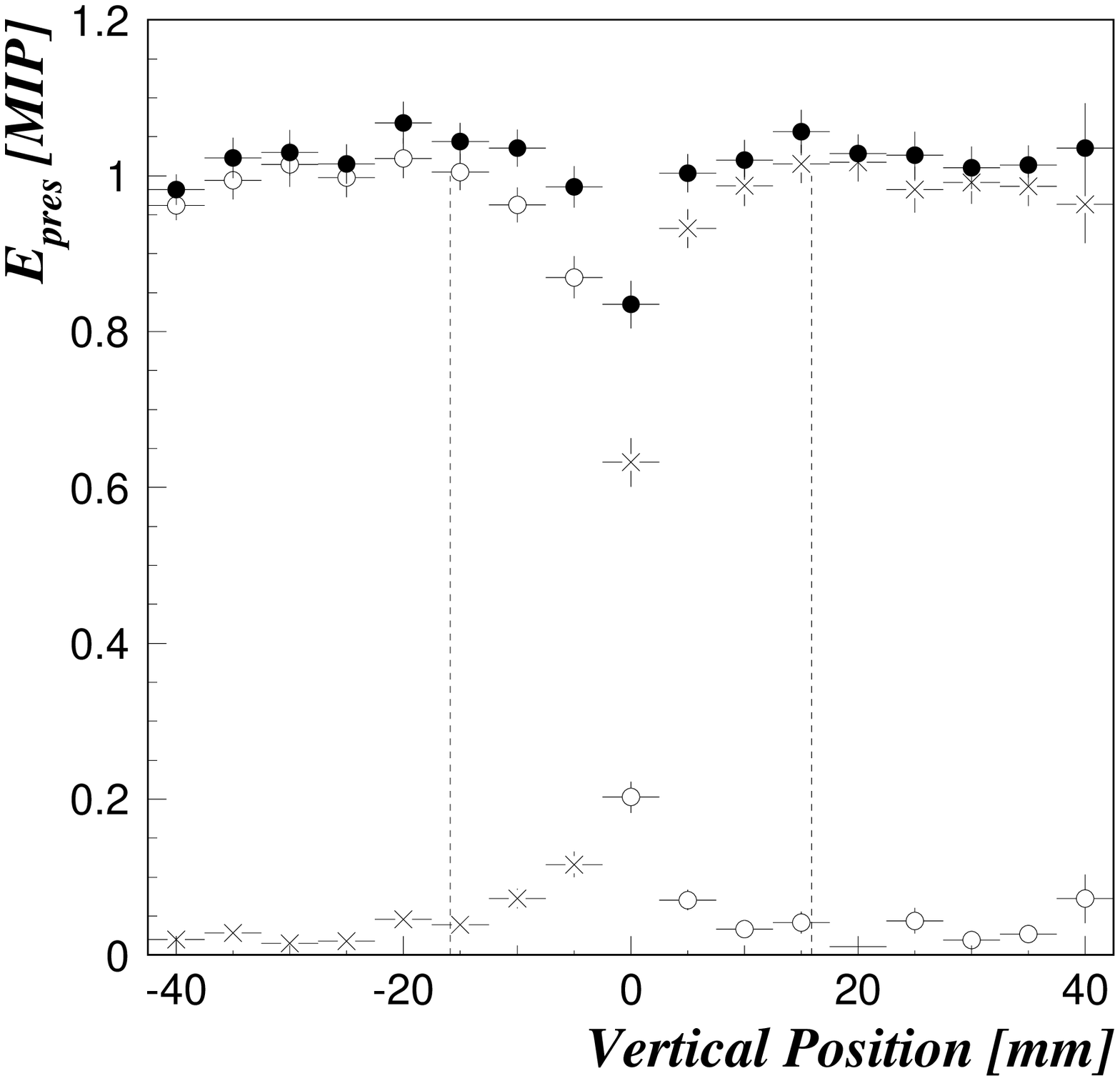,height=6cm}}
\put(10,134){\large{a)}}
\put(160,80){\large{b)}}
\put(320,80){\large{c)}}
\end{picture}
   \caption{\it 
The muon response uniformity of the presampler near tile borders.\hspace{2cm}
a) A sketch of scanning region,
b) a horizontal scan (* represents the response of
 tile 6, $\circ$ that of tile 7 and $\bullet$ the sum of both), 
c) a vertical scan, perpendicular to the embedded fibers.
The dashed lines indicates the fiber positions.
 The data in both plots are averaged over a uniformly populated rectangle
 90~x~5~mm$^2$ in area with the long side perpendicular to the scanning direction.}
   \label{f9-1}
\end{figure}

Figure~\ref{f9-1} shows the mean presampler response to 75 GeV muons.
The position information was provided by the delay wire
chamber. The presampler signals for the incident
muons are normalized
to the response at the center of the tile and averaged over
uniformly populated rectangles of 90~x~5~mm$^2$. The nonuniformity
in the sum of the two bordering tiles is a few percent in the regions
of the fibers. In the horizontal direction (figure~\ref{f9-1}b) the tiles
are mounted within one cassette with no gaps between them. 
In the vertical coordinate (figure~\ref{f9-1}c) 
a signal drop is observed between the cassettes 
due to the 1.4 mm gap between the scintillator tiles. 
The nonuniformity averaged over the surface of a tile is
less than 1\%.

\subsection{The presampler response to electrons}

The electron beam used for the energy correction studies was 1 cm wide and 10
cm high, centered horizontally within one FCAL module and vertically on
one of the 20~x~5~cm$^{2}$ electromagnetic sections.
The presampler signal ($E_{pres}$) was obtained by summing all 16 tiles in
order to be sure to get the entire signal and because the 
electronic noise contribution
was negligible. The signal from each tile was 
normalized to its average response to muons, resulting in units we refer to
as ``MIP''. Aluminium plates of 3 cm thickness were used as the absorber
material. In the following three such plates together are referred to as
one radiation length, an approximation which is accurate to 1\%.

\begin{figure}[tbp]
\centerline{\psfig{figure=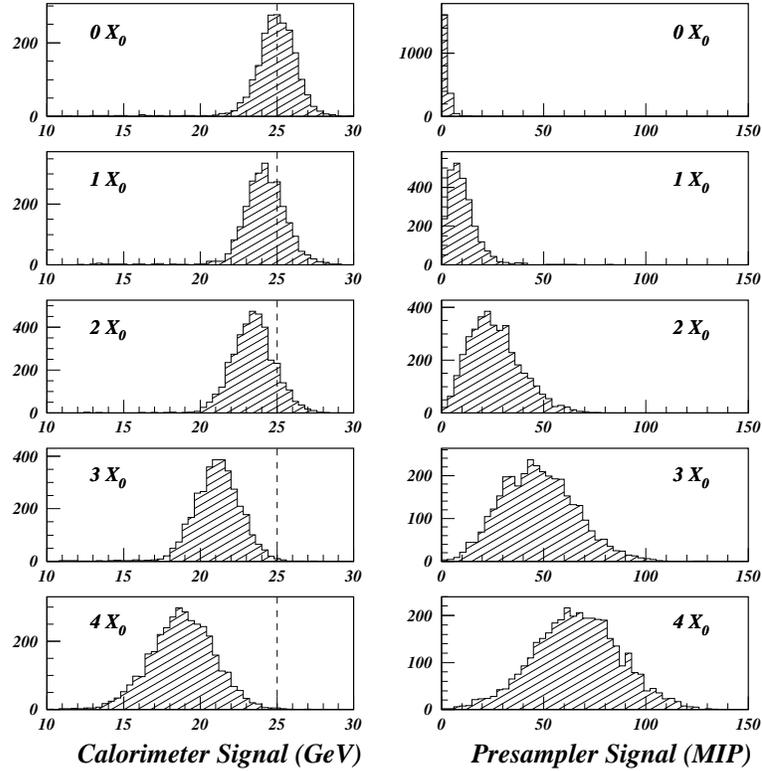,bbllx=10pt,bblly=130pt,bburx=550pt,bbury=700pt,height=12cm}}
  \caption{\it Signal distributions in the calorimeter and presampler for 25 GeV
electrons having passed through 0, 1, 2, 3, and 4 radiation lengths of 
aluminium absorber.}
   \label{calvpres}
\end{figure}

As examples of the calorimeter and presampler signal spectra
we show in figure~\ref{calvpres}
the energy distributions measured with the calorimeter ($E_{cal}$)
and the signal in the presampler for 25 GeV electrons for
aluminium absorber thicknesses ranging between 0 and 4 X$_0$.
The mean value for $E_{cal}$ decreases by more than 20\% but
the shapes of the distributions remain approximately gaussian. 
The resolution deteriorates substantially in the presence of more than
2 X$_0$ of absorber material.

\clearpage 

Table 1 shows the relative calorimeter signal loss and the average and RMS
values of the presampler signal spectra for the full range of electron
energies and absorber thicknesses.
The uncertainties presented are dominated by the statistical
precision.\\*[3mm]

\begin{tabular}{|c|c|rcl|rcl|rcl|}
\hline
Energy & Absorber & \multicolumn{3}{|c|}{Rel. Energy Loss} & \multicolumn{3}{|c|}{Presampler Avg} & \multicolumn{3}{|c|}{Presampler RMS} \\
(GeV) & ($X_0$) & \multicolumn{3}{|c|}{(\%)} & \multicolumn{3}{|c|}{(MIP)} & \multicolumn{3}{|c|}{(MIP)} \\
\hline
 3& 1& \hspace*{1mm}  7.8& $\pm$ &1.0&  5.5& $\pm$ &0.1& \hspace*{2mm}  3.7& $\pm$ &0.1\\
\cline{3-11}
       & 2& \hspace*{1mm} 18.0& $\pm$ &1.0& 10.3& $\pm$ &0.2& \hspace*{2mm}  5.3& $\pm$ &0.1\\
\cline{3-11}
       & 3& \hspace*{1mm} 32.8& $\pm$ &1.0& 12.0& $\pm$ &0.2& \hspace*{2mm}  5.6& $\pm$ &0.1\\
\cline{3-11}
       & 4& \hspace*{1mm} 49.4& $\pm$ &1.0& 12.1& $\pm$ &0.2& \hspace*{2mm}  5.5& $\pm$ &0.1\\
\hline
 5& 1& \hspace*{1mm}  6.4& $\pm$ &0.6&  6.2& $\pm$ &0.2& \hspace*{2mm}  4.1& $\pm$ &0.1\\
\cline{3-11}
       & 2& \hspace*{1mm} 13.0& $\pm$ &0.7& 12.8& $\pm$ &0.2& \hspace*{2mm}  6.5& $\pm$ &0.2\\
\cline{3-11}
       & 3& \hspace*{1mm} 27.4& $\pm$ &0.7& 16.9& $\pm$ &0.3& \hspace*{2mm}  7.3& $\pm$ &0.2\\
\cline{3-11}
       & 4& \hspace*{1mm} 42.5& $\pm$ &0.7& 19.5& $\pm$ &0.3& \hspace*{2mm}  7.1& $\pm$ &0.2\\
\hline
10& 1& \hspace*{1mm}  4.8& $\pm$ &0.5&  7.7& $\pm$ &0.2& \hspace*{2mm}  5.0& $\pm$ &0.1\\
\cline{3-11}
       & 2& \hspace*{1mm} 10.1& $\pm$ &0.5& 18.2& $\pm$ &0.4& \hspace*{2mm}  8.9& $\pm$ &0.3\\
\cline{3-11}
       & 3& \hspace*{1mm} 20.3& $\pm$ &0.5& 27.4& $\pm$ &0.4& \hspace*{2mm} 10.9& $\pm$ &0.3\\
\cline{3-11}
       & 4& \hspace*{1mm} 34.6& $\pm$ &0.7& 32.7& $\pm$ &0.6& \hspace*{2mm} 11.3& $\pm$ &0.4\\
\hline
15& 1& \hspace*{1mm}  3.5& $\pm$ &0.4&  7.9& $\pm$ &0.2& \hspace*{2mm}  5.2& $\pm$ &0.2\\
\cline{3-11}
       & 2& \hspace*{1mm}  7.8& $\pm$ &0.4& 20.4& $\pm$ &0.5& \hspace*{2mm}  9.6& $\pm$ &0.4\\
\cline{3-11}
       & 3& \hspace*{1mm} 17.3& $\pm$ &0.4& 34.9& $\pm$ &0.6& \hspace*{2mm} 13.8& $\pm$ &0.4\\
\cline{3-11}
       & 4& \hspace*{1mm} 29.6& $\pm$ &0.6& 46.0& $\pm$ &0.8& \hspace*{2mm} 14.6& $\pm$ &0.6\\
\hline
25& 1& \hspace*{1mm}  2.7& $\pm$ &0.3&  9.8& $\pm$ &0.2& \hspace*{2mm}  6.5& $\pm$ &0.2\\
\cline{3-11}
       & 2& \hspace*{1mm}  5.9& $\pm$ &0.3& 26.6& $\pm$ &0.4& \hspace*{2mm} 12.8& $\pm$ &0.3\\
\cline{3-11}
       & 3& \hspace*{1mm} 15.4& $\pm$ &0.3& 47.3& $\pm$ &0.5& \hspace*{2mm} 17.9& $\pm$ &0.4\\
\cline{3-11}
       & 4& \hspace*{1mm} 24.4& $\pm$ &0.3& 66.1& $\pm$ &0.6& \hspace*{2mm} 20.9& $\pm$ &0.4\\
\hline
50& 1& \hspace*{1mm}  1.6& $\pm$ &0.2& 12.0& $\pm$ &0.3& \hspace*{2mm}  7.9& $\pm$ &0.2\\
\cline{3-11}
       & 2& \hspace*{1mm}  4.6& $\pm$ &0.2& 35.6& $\pm$ &0.6& \hspace*{2mm} 16.4& $\pm$ &0.4\\
\cline{3-11}
       & 3& \hspace*{1mm} 11.5& $\pm$ &0.5& 72.7& $\pm$ &2.1& \hspace*{2mm} 27.7& $\pm$ &1.5\\
\cline{3-11}
       & 4& \hspace*{1mm} 19.8& $\pm$ &0.3&108.1& $\pm$ &1.2& \hspace*{2mm} 31.6& $\pm$ &0.8\\
\hline
\end{tabular}

\vskip 5mm

Table 1: {\it The relative decrease in the calorimeter signal and 
the average and RMS values of the presampler signal spectra for
each electron energy and each aluminium absorber thickness used 
in the test beam studies. The uncertainties shown are dominated by the
statistical precision.}

\clearpage

\subsubsection {Contribution of backscattering to the presampler signal}

\begin{figure}[tbp]
\centerline{\psfig{figure=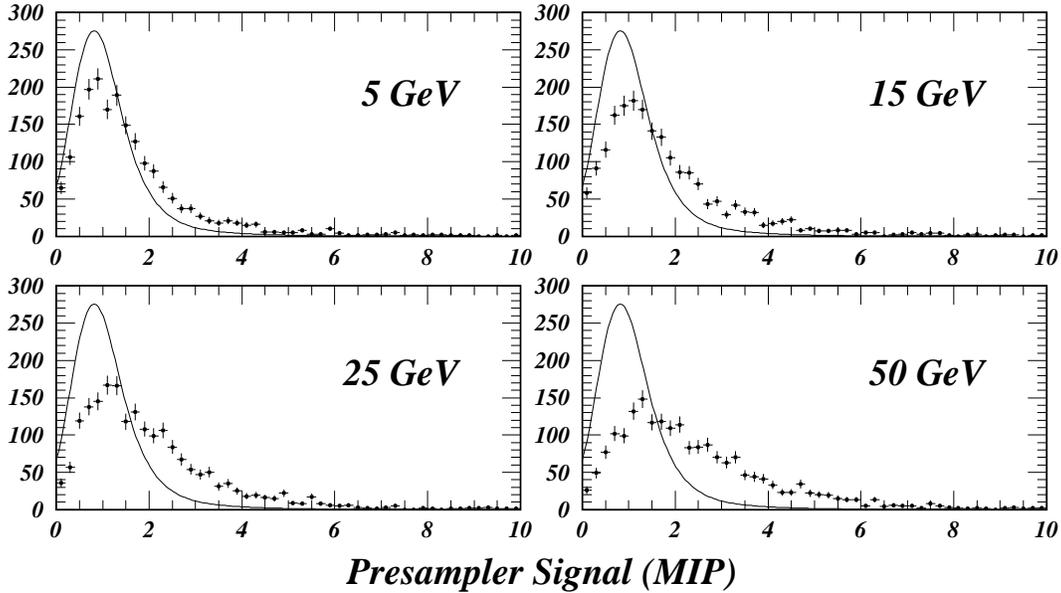,bbllx=50pt,bblly=270pt,bburx=530pt,bbury=540pt,height=8cm}}
 \caption{\it 
A comparison of the presampler signal spectra for 5, 15, 25, and 50 GeV
electrons to that for muons. The smooth curve shows the response to muons. 
The presampler
signal has been normalized to the average value of the muon spectrum. Each
spectrum contains 2000 entries.}

   \label{backscat}
\end{figure}

The comparison of presampler signal spectra from incident muons with those
of incident electrons allowed us to estimate the contribution of
backscattering from the electromagnetic showers in the uranium calorimeter.
Figure~\ref{backscat} shows this comparison for 5, 15, 25 and 50 GeV electrons.
 We determine
the average relative increase to be 1.45, 1.65, 1.80, and 2.16 respectively,
to an accuracy of 2\%. By adding absorber upstream of the presampler and
displacing both absorber and presampler several meters upstream to measure
the decreased contribution from backscattering, we ascertained that the
backscattering contribution is not increased by the presence of the absorber
in front of the presampler. Thus we can be sure that the backscattering
contribution remains at the level of 1 MIP and can be neglected at the level
of 10\% of the size of the signals we use for the electromagnetic energy
correction. We also measured the backscattering contribution from hadronic
showers and found for 15 and 75 GeV incident pions values for the average
relative increase less than 1.5 and 2.0 respectively.

\subsubsection{Electron energy correction}
\begin{figure}[tbp]
\centerline{\psfig{figure=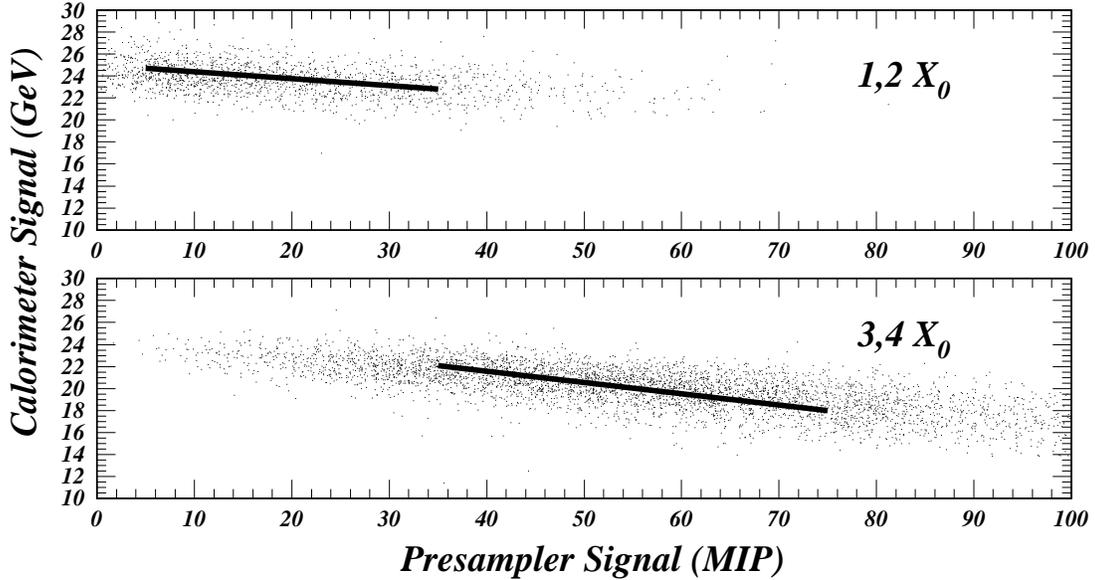,bbllx=25pt,bblly=270pt,bburx=535pt,bbury=545pt,height=8cm}}
   \caption{\it Calorimeter versus presampler response for 25 GeV electrons
 and absorber material ranging from 1 to 4 X$_0$. The line represents
the fit to the data according formula (1).}
   \label{f9-4}
\end{figure}

Figure~\ref{f9-4} shows the correlation between the presampler signal (normalized to
the average signal of a minimum--ionising particle) and 
the calorimeter
signal for 25 GeV electrons as a function of the amount of absorber material.

We have considered a variety of parametrisations for the relationship between
calorimeter and presampler responses. In the well--defined 
environment of a test beam
the correction is straightforward and depends on the incident
energy and the amount of absorber material, both of which are precisely known.

\begin{figure}[tbp]
\vskip -2cm
\centerline{\psfig{figure=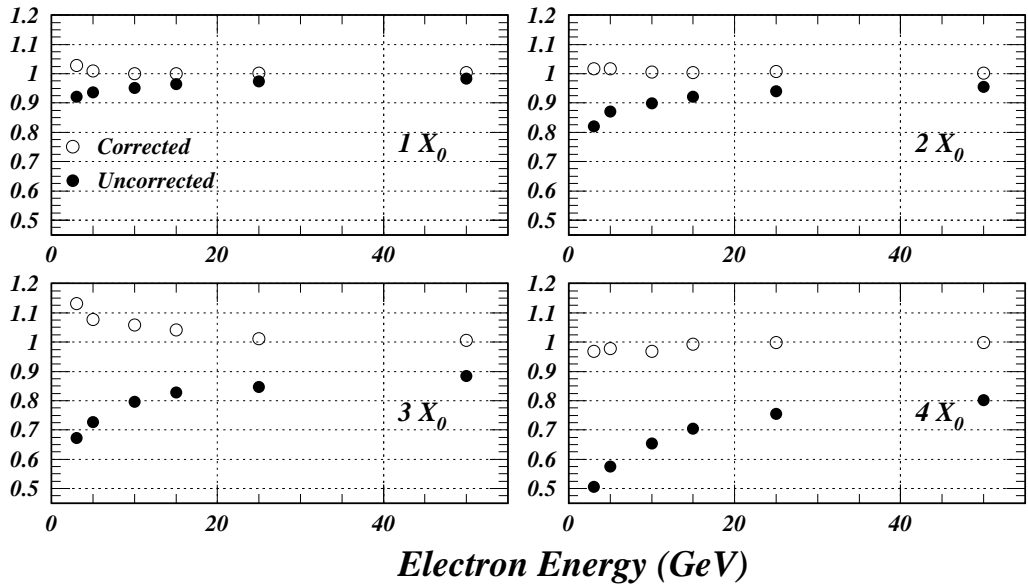,bbllx=50pt,bblly=265pt,bburx=530pt,bbury=540pt,height=8cm}}
   \caption{\it Average calorimeter response normalized to the
electron energy versus the electron energy before and after correction}
   \label{f9-2}
\end{figure}
In a detector environment the amount of absorber material in
front of the calorimeter is not uniformly distributed, arising from cables,
support structures, etc. One can, however, identify regions where the
average amount of absorber material is roughly known. For this reason
we show here the result obtained with 
one set of correction constants common to the 1 and 2 X$_0$
data set and one for the 3 and 4 X$_0$ data set.
The relation between the measured mean values of  $E_{cal}$ and $E_{pres}$
has been parametrised in a linear approximation:
\begin{equation}
\label{eq:etrue}
     E_{cal}= a_0 + a_1 E_{pres}
\end{equation}

The result for the two data sets for 25 GeV electrons is shown in figure~\ref{f9-4}.
The parameters $a_i$ depend on the amount of
material and on the electron beam energy.  This correction algorithm allows
for a linear energy dependence of the
parameters $a_i$: $a_i =
\alpha_i +\beta_i E_{beam}$. We neglect the dependence on the amount of
absorber material in order to estimate the success of the algorithm when
the amount of absorber varies within the data sample.
The parameters $\alpha_i$ and $\beta_i$ are determined by
minimising the difference of the beam energy and the 
corrected calorimeter signal,

The results
for the corrected calorimeter response for electrons in the energy
range 3-50 GeV, are
shown in figure~\ref{f9-2}.  This procedure provides a 
correction accurate to 3\% 
for the energy range studied here, but for an overcorrection
of about 10\% for the 3~X$_0$ data points at low energy.  
For electron energies greater than
5 GeV and for absorber thickness less than 2 X$_0$, the values relevant to
the operation of the ZEUS detector, this simple correction algorithm
yields a systematic precision of 2\%.

\begin{figure}[htbp]
\centerline{\psfig{figure=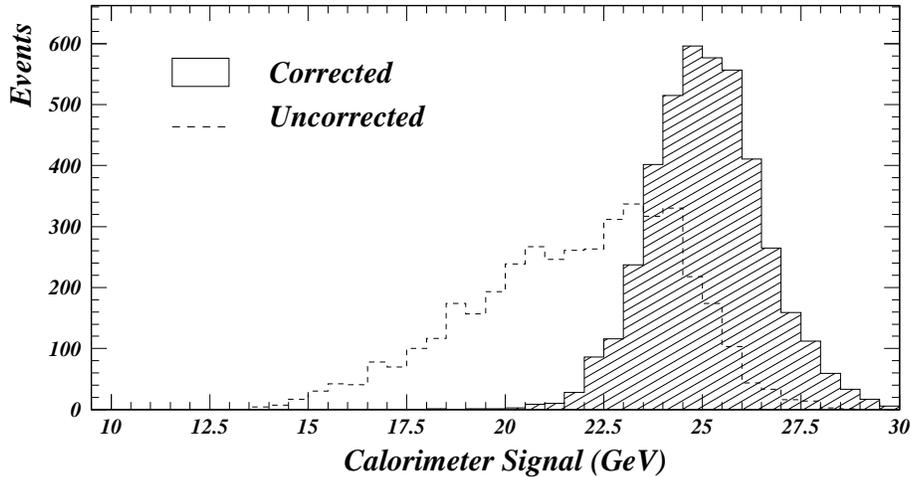,bbllx=25pt,bblly=270pt,bburx=530pt,bbury=540pt,height=6.5cm}}
   \caption{\it Reconstructed energy distributions for 25 GeV electrons for a mixture of the 1-4 X$_0$ data before and after application of the correction algorithm}
   \label{mixture}
\end{figure}

The improvement in the energy resolution as well as in the energy scale
is shown in figure~\ref{mixture}. The energy distribution
of 25 GeV electrons for a merged 1-4 X$_0$ data set is shown before
and after correction. 

\section {Summary and conclusions}

We have designed, built, installed and operated a scintillator-tile presampler
for the forward and rear calorimeter of the ZEUS detector. The scintillation
light from the tiles is collected by wavelength-shifting fibers and guided
to multi-anode photomultipliers via clear fibers. The signals are shaped, 
sampled and pipelined in the manner employed for the calorimeter itself,
easing the integration of the presampler in the ZEUS data acquisition system.
The performance of the tiles and fiber readout were monitored with a 
cosmic--ray telescope and with collimated sources during production.
The single-particle detection efficiency is greater than 99\% and the
response uniformity over the area of each of the 576 20~x~20~cm$^{2}$ tiles is 
better than 5\%. An LED flasher system and scans with radioactive sources
have proven useful diagnostic tools since the installation of the presampler. 
The in situ calibration with minimum--ionising particles during the 1995 
data--taking period achieved a precision better
than 5\% per tile. Test beam studies of a presampler prototype with the
final geometry and readout in combination with a prototype of the ZEUS
forward calorimeter verified the efficiency and uniformity results
and allowed the determination of backscattering contributions. Tests
with electrons were performed in the energy range 3--50 GeV with
0--4 radiation lengths of aluminium absorber. These studies
have proven the feasibility of an electron energy
correction accurate to better than 2\% in the energy range and for the
configuration of inactive material relevant to the ZEUS detector.

\section {Acknowledgements}

We would like to thank the technical support from the institutes 
which collaborated
on the construction of the presampler, 
in particular W. Hain, J. Hauschildt, K. Loeff\-ler, A. Maniatis, 
H.--J. Schirrmacher (DESY), 
W. Bienge, P. Pohl, K.--H. Sulanke (DESY--IfH Zeuthen),
M. Gospic, H. Groenstege, H. de Groot, J. Homma, I. Weverling, 
P. Rewiersma (NIKHEF), R. Mohrmann, H. Pause, W. Grell (Hamburg), 
M. Riera (Madrid), R. Granitzny, H.--J. Liers (Bonn). We are also grateful for the 
hospitality and
support of CERN, in particular the 
help of L. Gatignon is much appreciated. 
Finally we would like to thank R. Klanner for his enthusiastic support during all
phases of the project.

\end{document}